\documentclass[authoryear,longtitle,12pt]{elsarticle}
\usepackage{setspace} 
\usepackage{amsmath}
\usepackage{ulem}
\usepackage{graphicx}
\usepackage[margin=1ex]{subfig}
\usepackage{natbib}
\usepackage{amsmath}
\usepackage{amsfonts}
\usepackage{amssymb}

\parskip=\baselineskip

\newcommand{\ci} {\textbf{citation needed\ }}
\newcommand{\hide}[1] {}

\newcommand{\apislong}{\textit{Apis mellifera}}
\newcommand{\apis}{\textit{A. mellifera}}

\newcommand{\note}[1]{\textit{[ #1 ]}}
\newcommand{\dt}[1]{\frac{d #1}{d t}}
\newcommand{\changes}[1]{\textbf{*** #1 ***}}

\begin{document}

\begin{frontmatter}
\title{Intelligent Decisions from the Hive Mind: \\ Foragers and Nectar Receivers of \textit{Apis mellifera} Collaborate to Optimise Active Forager Numbers}
\author[rvt]{James R. Edwards\corref{cor1}}
\ead{James.Edwards@sydney.edu.au}
\author[rvt]{Mary R. Myerscough}
\address[rvt]{The School of Mathematics and Statistics\\
and The Centre for Mathematical Biology\\
University of Sydney, New South Wales, 2006, Australia.}
\cortext[cor1]{Corresponding author}
\begin{abstract}
\hide{Honey bee foraging is a well-known example of collective decision making. Foraging is an expensive investment by a colony; energy is expended in flying to the nectar source and there is a high risk of death to the forager which diminishes future foraging capability. For such a critical task performed by such an elegantly complex species it should be expected that \apislong\ has evolved techniques to balance the efficient exploitation of forage sources with the preservation of their limited population of foragers. Rather than being a collection of homogeneous individuals, the closely related bees that make up colonies of \apis\ can be categorised by behaviour. }We present a differential equation-based mathematical model of nectar foraging by the honey bee \apislong. The model focuses on two behavioural classes; nectar foragers and nectar receivers. Results generated from the model are used to demonstrate how different classes within a collective can collaborate to combine information and produce finely tuned decisions through simple interactions. In particular we show the importance of the `search time' - the time a returning forager takes to find an available nectar receiver - in restricting the forager population to a level consistent with colony-wide needs.
\end{abstract}
\end{frontmatter}

\section{Introduction}

Collective decisions may result from the simple aggregation of individual preferences. Often, however, they emerge as individuals' preferences and their interactions with others combine in complex ways. In some important situations, individuals can be categorised by their behaviour into different classes whose interactions drive the decision making process. Such scenarios occur in both human and animal societies \citep{conradLiszt200}. For example, social choice theory \citep{arrow1963} and its related fields \citep{dryzek2003Soci} are primarily concerned with how human societies make decisions when classes of individuals assign different utility measures to the available options. Examples in animal societies are given by \cite{couzin2005} and \cite{raghib2010Adve}, who divide social groups such as fish, insects and birds into classes of informed and uninformed individuals which each move according to different rules whilst interacting to ensure the group travels cohesively, and by \cite{cluttonBrock19} who discusses how groups decide which members will reproduce in societies containing dominant and subordinate females.\hide{also Escudero 2010}

\begin{figure}[htbp]
  \centering
  \includegraphics[]{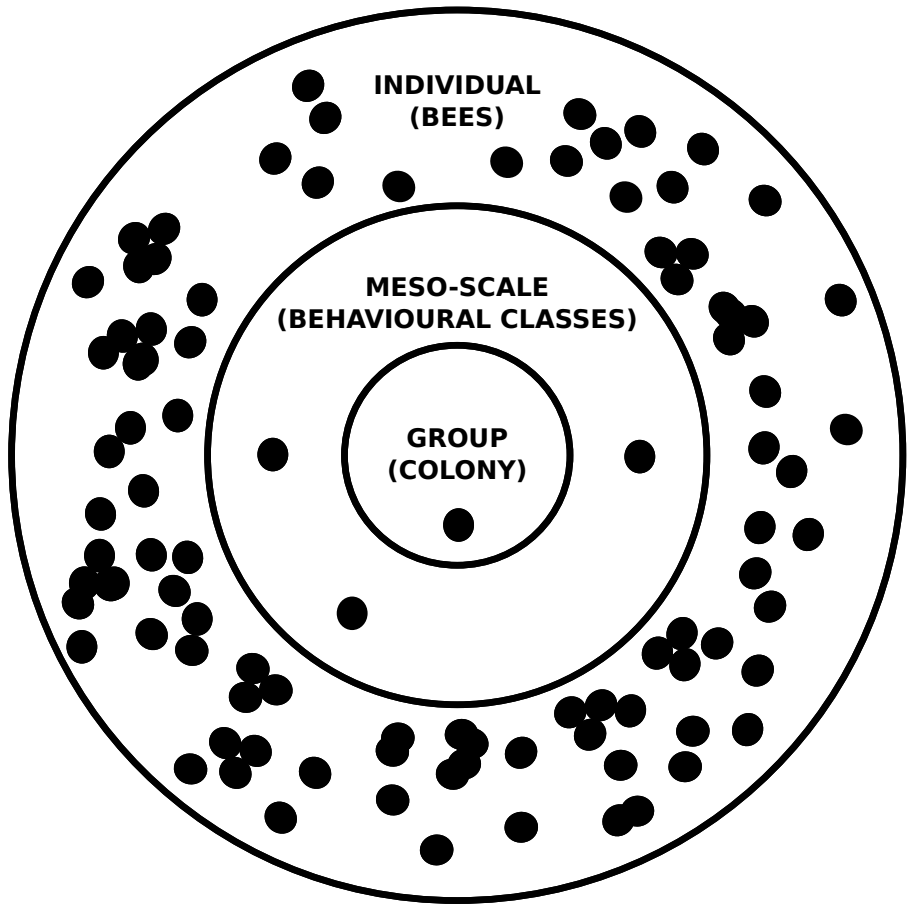}
  \caption{The level of organisation at which a collective decision is formed may occur at different scales. In some cases considering the group as a single entity is appropriate (inner circle), in other cases focusing on each of the potentially thousands or millions of individuals is more natural (outer circle). For nectar foraging by colonies of \apis\ modelling interaction at the mesoscale (middle circle), that is between a few different behavioural classes, allows us to best understand the forces driving the decision making process.}
  \label{fig:scales}
\end{figure}

Most collective decision models focus on either the individual-scale \citep{bonabeau2002abm,pratt2005abm} or the group-scale \citep{nevai2009,tachikawa2010O}, and therefore do not explicitly include interactions between classes which occur at an intermediate level of organisation. In this paper we look at the intermediate, or mesoscale, interactions between two behavioural classes of \apislong\ (honeybees), foragers and receivers, and explore how this interaction mediates the decision `how many foragers should we send out of the hive to gather nectar?' A representation of these three scales, and how they relate to each other, is given in Fig. \ref{fig:scales}.

\hide{Superorganisms, which are a ``collection of agents which can act in concert to produce phenomena governed by the collective" \citep{kelly1995out}, are often used to study collective decision making. We seek a model system that can be analysed mathematically to derive greater understanding of this type of problem. One quintessential superorganism, honey bees, \apislong, progress through different behavioural classes as they develop and are required by their colony in response to internal and external conditions.}

Nectar foraging is a particularly important and well studied task of the honey bee colony \citep{lindauer1961,seeley:wh,dyer2002bdl}. Colonies require energy in the form of nectar to survive and thrive. Foragers exploit nectar sources (flowers) by collecting their nectar and returning it to the hive. The nectar is delivered to waiting receiver bees who then process and store the nectar within the hive. The forager's main behavioural choices are resting, returning (to the forager source) or recruiting (performing the waggle dance to encourage other resting foragers gather nectar)  \citep{camazineSneyd1,dyer2002bdl}.

Foraging is not only important; it is dangerous and costly. There is a high risk to the forager of predation and misadventure resulting in death \citep{dukas2001epd,page2001aad}. A forager expends energy when she flies out on a foraging expedition, which may be unsuccessful. Both forager death and unsuccessful foraging may diminish nectar inflow and, if frequent, potentially lead to a rapid decline in the colony population (Khoury et al. 2010) \note{submitted - will update upon acceptance or rejection.}. Clearly, if it is to prosper, the colony must make a wise collective decision about how best to allocate foragers for maximum nectar return. The principle motivation of this paper is to model the effect on forager recruitment of forager and receiver interactions, and thereby understand more about their impact on the forager allocation decision.

Existing models of nectar foraging focus, unsurprisingly, on foragers \citep{camazineSneyd1,devries1998mcf,cox2003fmf}. In particular, they focus on the recruitment of new foragers through the waggle dance and the communication of nectar source quality to other foragers. All such models relate recruitment rates to nectar source quality; under normal conditions high quality sites will induce extensive waggle dancing and high recruitment levels whereas low quality ones may not even lead to waggle dancing at all. These models explain how a colony allocates foragers in accordance with the quality of the environment, but they do not consider the extent of the colony's need for nectar. Through our model we seek to explain how, by balancing quality and need, the colony makes a decision that will best ensure its future survival.

Experienced foragers know about the quality of the foraging environment, which includes not only the concentration of sugar in the nectar but also all other factors that a forager evaluates when determining the number of waggles runs they will perform, but they know little about the state of the hive's nectar reserves. Conversely, receivers know little of the quality of the foraging environment but learn about the state of the colony's reserves as they store nectar \citep{seeley1996hbs}. We introduce a model that incorporates both foragers (knowledgeable about the external environment) and receivers (knowledgeable about the internal environment) to show how interactions between these classes determine the size of the active forager population. The model uses a system of three nonlinear ordinary differential equations representing the populations of active foragers, foragers unloading their nectar, and receivers available to unload nectar. We find both analytic and numerical results of the model and use these results to draw conclusions about \apis\ foraging in particular and collective decision making in general. Finally, we extend the model to incorporate multiple, dynamic foraging sources and changes in receiver population. These models are neither individual-oriented nor group-oriented. They operate at the mesoscale (Fig. \ref{fig:scales}) in which classes of individuals (in this model foragers and receivers) are the entities that are considered.

\hide{In considering a decision, collective or otherwise, we need to identify the stimuli, the evaluation process and the actual decision outcome \citep{brim1978pdp,hansson1994dt} (figure ...). We can consider two stimuli: forage source quality and colony nectar need. We also know the potential outcomes (the choices that we are interested in): they are the set of the possible populations of active foragers over the duration of foraging activity. The model that we construct will allow us to explore the evaluation process, that is, how the colony as a group `decides' which of the possible decision outcomes to choose on the basis of the stimuli.

We use this model to demonstrate how the intelligent collective decision about the ideal size of the active forager population emerges from interactions between individuals belonging to different behavioural classes. This is first model to incorporate the knowledge of receivers directly into the forager recruitment process, and the first ODE model in which the conclusions of \cite{seeley1994stf} have been applied (although the individually oriented model of \cite{devries1998mcf} makes the individual forager's likelihood to dance a function of search time \note{search time needs to be defined}). The results of our model demonstrate how the interaction of foragers and receivers shapes the population of active foragers in response to forage quality and colony need rather than the exhaustion of the available forager population \citep{cox2003fmf} or duration of foraging activity \citep{camazineSneyd1}.}

\section{Simple Model\label{sec:simpleModel}}

We consider two behavioural classes of \apis, foragers and receivers, and construct a model to explore how the interaction of these classes determines the active forager population. We assume that the foragers' tendency to rest, to return or to recruit is determined by two factors;\hide{genetically determined individual preference oldroyd1993gvh,} the quality of the forage source \citep{camazineSneydS} and the time spent searching for a receiver \citep{seeley1994stf}. Initially, we introduce two simplifying assumptions; the total receiver population is considered to be constant throughout the model\hide{(though this recognises that some of these bees may be performing other roles until needed as receivers)} and there is a single nectar source of constant quality. Later we will extend the model and drop these assumptions.

\begin{figure}[htbp]
  \centering
  \includegraphics[width=0.5\paperwidth]{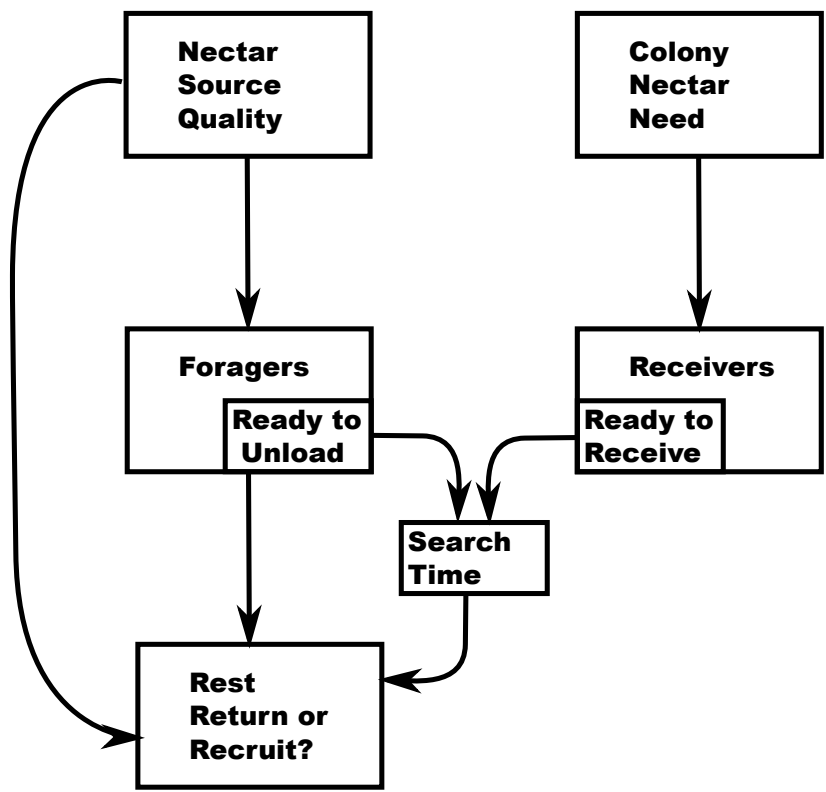}
  \caption{The decision making process of a colony of \apislong\ engaged in nectar foraging. Nectar source quality and colony nectar need are the two inputs, perceived and analysed by foragers and receivers respectively. The interactions of foragers and receivers produce a value, search time. Nectar source quality and search time are used by foragers to decide whether to rest, return or recruit.}
  \label{fig:decisions}
\end{figure}

The interactions in this model are illustrated in Fig. \ref{fig:decisions}. Two separate sets of data are perceived by different classes of the colony, nectar source quality by the forager class and colony nectar need by the receiver class. The colony nectar need determines the population of receivers and the `ready to receive' subset of this population. This population, and the population of those foragers ready to unload, determines the average time that a forager spends searching for a receiver to unload nectar, known as the `search time'. A forager's tendency to choose to rest, to return or to recruit is dependent upon the nectar source quality and the search time \citep{seeley1994stf}. The outcome of each individual decision then feeds back into the population of foragers, and hence into future search times. This system places search time at the centre of a web of interactions, and so we will describe and define it in more detail.

\subsection{Search time of returning foragers for receivers \label{sec:searchTime}}
\hide{\cite{seeley1994stf} proposed an `urn model' to estimate the expected time it takes a newly returned forager to find a receiver who will accept their load of nectar, we follow this model closely. It is important to note that this provides an average value, in practise there is variability in individual search times.} Upon entering the hive after a successful foraging expedition, a returning forager seeks a receiver who will accept her load of nectar. The expected time spent searching for a receiver can be modelled using the proportion of foragers and receivers \citep{seeley1994stf}. Let $F$ be the total active forager population, $R$ be the total receiver population, $\bar{F}$ be the population of foragers seeking to unload near the hive entrance and $\bar{R}$ be the population of receivers ready to receive. Then, assuming that the population in the hive entrance is exclusively comprised of foragers seeking to unload and receivers willing to receive nectar loads, the probability that a forager's approach to another bee in the entrance is to a receiver can be modelled by $\bar{R} / (\bar{F} + \bar{R})$. If we now let $s_s$ be the constant time taken when approaching another bee then the expected search time at time $t$, $S(t)$, is 
\begin{equation}
S(t) = s_s (\bar{F} (t) + \bar{R}(t))  / \bar{R} (t).
\label{eq:searchTime}
\end{equation}
The expected search time $S(t)$ is critical; it aggregates the outcomes of numerous interactions between individual foragers and receivers into a single value that strongly influences the rate of forager recruitment in the model. It is worth noting that this model, because it is a system of ordinary differential equations, can only be used to calculate the average search time when, in reality, the actual search time experienced by individual foragers will vary significantly around this average. Nevertheless, since we are modelling a colony wide phenomenon rather than individual behaviour, using the average search time still gives a reasonable indication of the expected behaviour of the system.

\subsection{Equations for forager and receiver numbers}
\begin{figure}[htbp]
 \centering
  \subfloat[\label{fig:gamma1s}]{\includegraphics{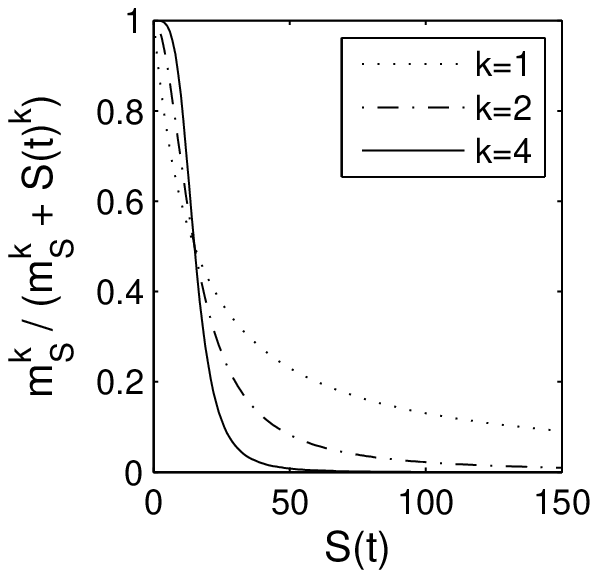}}
  \subfloat[\label{fig:gamma1q}]{\includegraphics{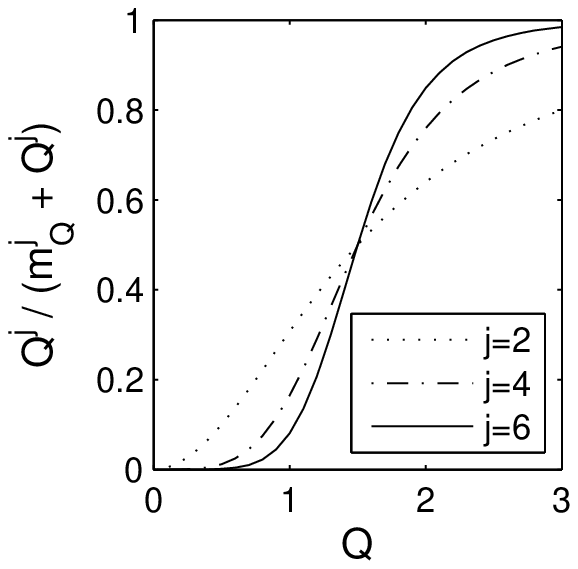}}
 \caption{Plots of the Hill equation terms in $\Gamma_1$ with different coefficient values. Half-maximal values are derived by comparison with experimental data \citep{seeley1992Trem,seeley:wh}. Coefficient values are chosen by comparing these curves to this experimental data. Fig. \ref{fig:gamma1s} shows how increases in search time decrease recruitment. For very low search times, recruitment (effectively the number of waggle runs the forager performs due to quality) is essentially unchanged. For very high search times waggle dancing may effectively cease. Three different curves are shown representing different values of the coefficient $k$. Fig. \ref{fig:gamma1q} shows how increases in quality lead to foragers recruiting more. For low quality sources recruitment is negligible. As the quality approaches $1.5$, recruitment is half the maximal rate. As quality approaches $3$ the rate of recruitment plateaus. Three different curves are shown representing different values of the coefficient $j$.}
\label{fig:gammaPlots}
\end{figure}

Let $Q$ be the quality of a nectar source. Here, quality incorporates all criteria that foragers assess when determining their likelihood to waggle dance. These criteria might include things like risk of predation \citep{dukas2001epd} but the most influential component is likely to be sugar concentration. Foragers exploiting this source are increasingly likely to recruit (i.e. will perform more waggle runs) as $Q$ increases or as $S(t)$ decreases. Conversely, their likelihood of resting increases for low $Q$ and for high $S(t)$. Defining $f_r$ and $f_s$ as maximal forager recruitment and resting rates respectively we construct an equation for the changes in forager population, $F(t)$,
\begin{equation}
\frac{d F(t)}{dt} = \underbrace{f_r F(t) \ \Gamma_1 (S(t),Q)}_{\text{recruitment}}  - \underbrace{f_s F(t) \  \Gamma_2 (S(t),Q).}_{\text{resting}} \label{eq:sysF}
\end{equation}
Here $\Gamma_1$ and $\Gamma_2$ are functions that scale the recruitment and resting rates by mapping source quality and search time to values between 0 and 1. $\Gamma_1$ should increase recruitment rates as quality improves or as search time decreases. $\Gamma_2$ should increase resting rates as quality decreases or as search time increases. We have chosen to define these functions using Hill equations since this type of equation permits a good phenomenological fit and is algebraically convenient. We define $m_Q$ and $m_S$ as the values that produce half-maximal scalings for $Q$ and $S(t)$, and $j$ and $k$ as the coefficients that control the sensitivity of the functions to changes in $Q$ and $S(t)$. This leads to definitions for $\Gamma_1$ and $\Gamma_2$,
\begin{align}
\Gamma_1(S(t),Q) = & \frac{m_S^k}{S(t)^k + m_S^k} \frac{Q^j}{Q^j + m_Q^j} 	\label{eq:gamma1} \\
\Gamma_2(S(t),Q) = & \frac{S(t)^k}{S(t)^k + m_S^k} \frac{m_Q^j}{Q^j + m_Q^j}.	\label{eq:gamma2} 
\end{align}
In Fig. \ref{fig:gamma1s} we show how different search times affect the search-related part of $\Gamma_1$ for a range of coefficient values. Compared to foragers, there is a paucity of information about receivers in the existing literature and so we rely extensively on observations recorded by \cite{seeley1992Trem,seeley:wh}. We find that the closest comparison to the data given in \citet[Fig. 7]{seeley1992Trem} is with $k=4$ and $m_S=10$. Similarly, we plot the quality term of $\Gamma_1$ for a range of coefficients in Fig. \ref{fig:gamma1q} and determine that the best fit to the data in Fig. 5.31 of \cite{seeley:wh} is $j=4$ and $m_Q=1.5$. In the absence of similar data for resting rates we used these same coefficients for the equivalent terms in $\Gamma_2$. This also keeps the number of parameters in the model to a minimum. Eq. \ref{eq:sysF} emphasises nectar source quality and search time, it does not consider other factors such as the tendency of foragers to resume foraging at sources they were exploiting the previous day.

Next we define the changes in the population of foragers that are seeking to unload nectar within the hive. These foragers, denoted $\bar{F}$, are a subset of the total foragers $F$. The number of foragers not seeking to unload, i.e. those outside the hive gathering nectar, is $F(t)-\bar{F}(t)$. These foragers arrive at the hive with nectar at a rate determined by the average foraging trip time $f_a$. They then take an average of $S(t)$ seconds to find a receiver and commence unloading. The equation for this population is
\begin{align}
\frac{d \bar{F}(t)} {d t} = \underbrace{\frac{F(t)-\bar{F}(t)}{f_a}}_{\text{rate of arriving in entrance}} -\underbrace{\frac{\bar{F}(t)} {S(t)}.}_{\text{rate of finding a receiver}} \label{eq:sysFb}
\end{align}
For this simple model we assume that $R(t)$, the total number of available receivers, does not change and so $R(t) = R_0$, the initial receiver population. Each receiver can be either waiting to receive nectar or engaged in nectar storage tasks. The population of receivers engaged in storing is given by $R(t) - \bar{R}(t)$. They complete this task at a rate determined by the average time for nectar storage, denoted $r_s$. Meanwhile, as foragers find receivers available for unloading, these receivers drop out of the available receivers population and join those storing nectar. The equation for the population of available receivers is
\begin{align}
\frac{d \bar{R}(t)} {d t} = \underbrace{\frac{R_0 - \bar{R}(t)} {r_s}}_{\text{rate of returning from storing}} - \underbrace{\frac{\bar{F}(t)} {S(t)}.}_{\text{rate of receiving nectar}} \label{eq:sysRb}
\end{align}
The complete system of ODEs from Eqs \ref{eq:sysF}, \ref{eq:gamma1}, \ref{eq:gamma2}, \ref{eq:sysFb} and \ref{eq:sysRb} is
  \begin{align}
  \frac{d F(t)}{dt} = & f_r F(t)    \frac{m_S^k}{S(t)^k + m_S^k} \frac{Q^j}{Q^j + m_Q^j}                   - f_s F(t)            \frac{S(t)^k}{S(t)^k + m_S^k} \frac{m_Q^j}{Q^j + m_Q^j}  \notag \\
  \frac{d \bar{F}(t)} {d t} = & \frac{F(t)-\bar{F}(t)}{f_a} - \frac{\bar{F}(t)} {S(t)}  \label{eq:system}\\
  \frac{d \bar{R}(t)} {d t} = & \frac{R_0 - \bar{R}(t)} {r_s} - \frac{\bar{F}(t)} {S(t)}. 	\notag
  \end{align}
Unless otherwise stated, we will use the parameters given in Table \ref{tab:param}. In many cases a single value has been chosen from a range of appropriate values (e.g. the average forager trip time varies greatly).
\begin{table}[hbp]
\begin{scriptsize}
\begin{tabular}{llll}
\textit{Parameter}&\textit{Symbol}&\textit{Value}&\textit{Source}\\
\hline
Single interaction time			&$s_s$	& $5 \ s$ 		&	\cite{camazineSneydS} \\
										&			& (range is $2 \ s \to 7 \ s$) 	&	 \\
Forager recruitment rate		&$f_r$	& $0.0010	\ s^{-1}$ 		& 	\cite{seeley1992tdc} 	\\
Forager resting rate				&$f_s$	& $0.0002 	\ s^{-1}$	& 	\cite{camazineSneyd1}	\\
Half-maximal search time		&$m_S$	& $10			\ s$		&	\cite{seeley1992Trem} \\
Search time coefficient			&$k$		& $4$			&	\cite{seeley1992Trem} \\
Half-maximal forage quality	&$m_Q$	& $1.5 $		&	\cite{seeley1986sfh} \\
Forage quality coefficient		&$j$		& $4$			&	\cite{seeley1986sfh} \\
Forager round-trip time					&$f_a$	& $15 \ min$ 		&	\cite{seeley1994hbf} \\
										&	& (range is $1/2 \ min \to 20 min$) 	&	 \\
Receiver storage time			&$r_s$	& $20 \ min$		&	\cite{seeley:wh}	\\
					&	& (range is $1 \ min \to 20 min$) 	&	 \\
\hline
\end{tabular}
\caption{Standard parameter values}
\label{tab:param}
\end{scriptsize}
\end{table}
\subsection{Steady states\label{sec:steadyStates}}

We seek information about the structure and stability of the system and so consider the equilibrium points $(F^*, \bar{F}^*, \bar{R}^*)$ of the simple model (Eq. \ref{eq:system}). This system has three equilibrium points. The first corresponds to the state observed at the start of the day in which no foragers are active, the second to the population mix that the colony approaches over the course of the day in a normal foraging environment and the third to a biologically impossible situation where the population of foragers seeking to unload is negative.\hide{ Details of the derivation, stability and definitions of these equilibrium points can be found in the appendix.} Here we will restrict our attention to the first two equilibria, the one near which the colony typically begins and the one which it typically approaches during the course of a day.

The first equilibrium point is 
\[ 
\left( \begin{array}{c}
F^* \\
\bar{F}^* \\
\bar{R}^*
 \end{array} \right)
 = \left( \begin{array}{c}
0 \\
0 \\
R_0
 \end{array} \right)\] 
in which there are no foragers active and no receivers storing nectar. To find when the forager population increases away from this point we determine when it is unstable. We find \hide{(see appendix) }that this steady state is unstable whenever \hide{$Q > m_Q \sqrt[j]{\frac{f_s s_s^k} {f_r m_S^k}}$. }$Q > m_Q (f_s / f_r)^{1/j} (s_s / m_S)^{k/j}$. For the parameter values of Table \ref{tab:param} this corresponds to $Q \gtrsim 0.5$, so for all except very low quality environments the forager populations will increase from zero.

\begin{figure}[htbp]
  \centering
  \includegraphics[]{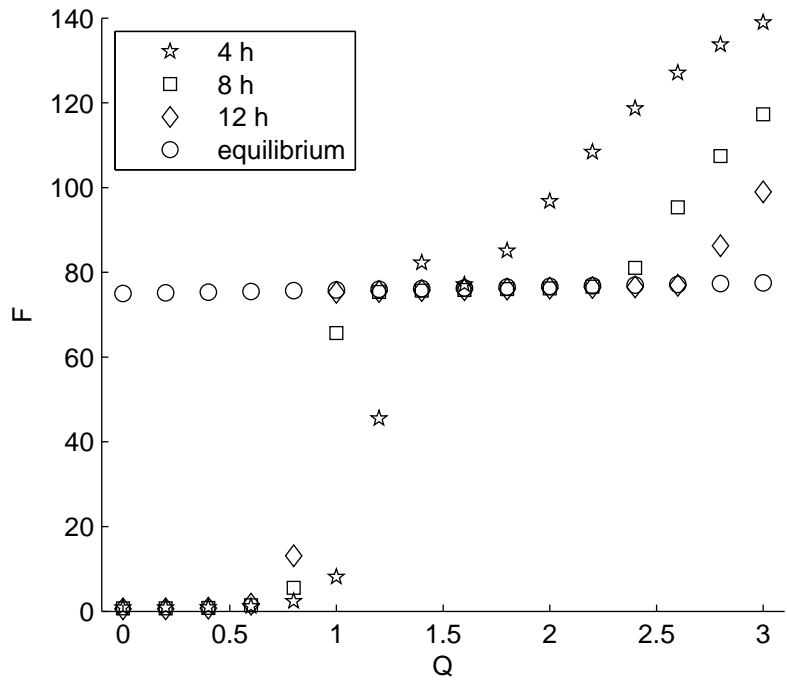}
  \caption{The population of foragers after 4, 8 and 12 hours for simulations with different values of $Q$ and otherwise identical parameters. Actual populations are shown as stars, squares and diamonds, the equilibrium population as a circle. The equilibrium point changes only slightly as quality increases as it is principally dependent upon $R$. For low values of $Q$ (i.e. $Q<1$), the population does not grow close to equilibrium, even after twelve hours. For moderate values of $Q$ (i.e. $Q \in [1,2]$), the population approaches equilibrium quickly and smoothly. For high values of $Q$ (i.e. $Q>2$) the population rapidly exceeds the equilibrium population and then, upon entering a stabilising phase dominated by inhibition of recruitment, slowly decreases towards it.}
  \label{fig:bar}
\end{figure}

The second equilibrium point is
\[ 
\left( \begin{array}{c}
F^* \\
\bar{F}^* \\
\bar{R}^*
 \end{array} \right)
 = R_0 \left( \begin{array}{c}
\frac{1}{r_s} \left( f_a + m_S \gamma \right) \\
 \left( \frac{s_s m_S \gamma} {(r_s +s_s) m_S \gamma - r_s s_s} \right) \left( \frac{m_S}{s_s} \gamma	-1 \right) \\
\frac{s_s m_S \gamma} {(r_s +s_s) m_S \gamma - r_s s_s}
 \end{array} \right)\]  
for $\gamma = \sqrt[k]{\frac{f_r}{f_s} Q^j m_Q^{-j} }$. This shows that the final populations are linearly related to the total receiver population, and are only slightly dependent upon source quality (through $\gamma$). In other words, the population of receivers determines how many foragers should be allocated. Calculating the stability of this point is analytically intractable, but numerical simulations indicate it is an attracting stable point and an analytic analysis of an approximation to it, i.e. of
\begin{equation}
\left( \begin{array}{c}
F^* \\
\bar{F}^* \\
\bar{R}^*
 \end{array} \right)
 \approx \frac{R_0}{r_s} \left( \begin{array}{c}
f_a \\
m_S\gamma \\
s_s
 \end{array} \right)
 \label{eq:stableApprox} 
 \end{equation}
\hide{which is stable, i.e. nearby points are attracted towards it. See Sec 4.3 of Glenndinning for why this works - i.e. for hyperbolic stationary points their stability is robust to small perturbations}
also suggests it is always stable. This analysis of the system dynamics indicates that in a normal foraging environment the forager population will increase from zero towards a single stable value. It is surprising that this value is principally dependent on forager arrival rate, receiver storage time and receiver population. In some cases, the stable equilibrium point in the model is approached but not reached within the limited time that a colony has to forage each day. We note that for low search times, $\frac{d}{d t} F(t) \approx \frac{f_r Q^j - f_s m_Q^j} {Q^j + m_Q^j} F(t)$, so that $F(t) \approx F_0 exp(\frac{f_r Q^j - f_s m_Q^j} {Q^j + m_Q^j} t)$, where $F_0$ is the initial population. Clearly, the population grows faster for higher values of $Q$ than for low values of $Q$.  So $Q$ strongly determines the \textit{rate} that the colony approaches the steady state (in the model) but only weakly affects the \textit{value} of that steady state. This is illustrated for the parameters of Table \ref{tab:param} in Fig. \ref{fig:bar}. This suggests that the forager population is more likely to attain (and even, temporarily, exceed) the stable equilibrium point in a high quality environment than in a low quality one. To understand the details of how this works in practice, we solve the system numerically.
\section{Numerical Results\label{sec:numerics}}

\begin{figure}[htbp]
 \centering
  \subfloat[\label{fig:timecourseHQHR}]{\includegraphics{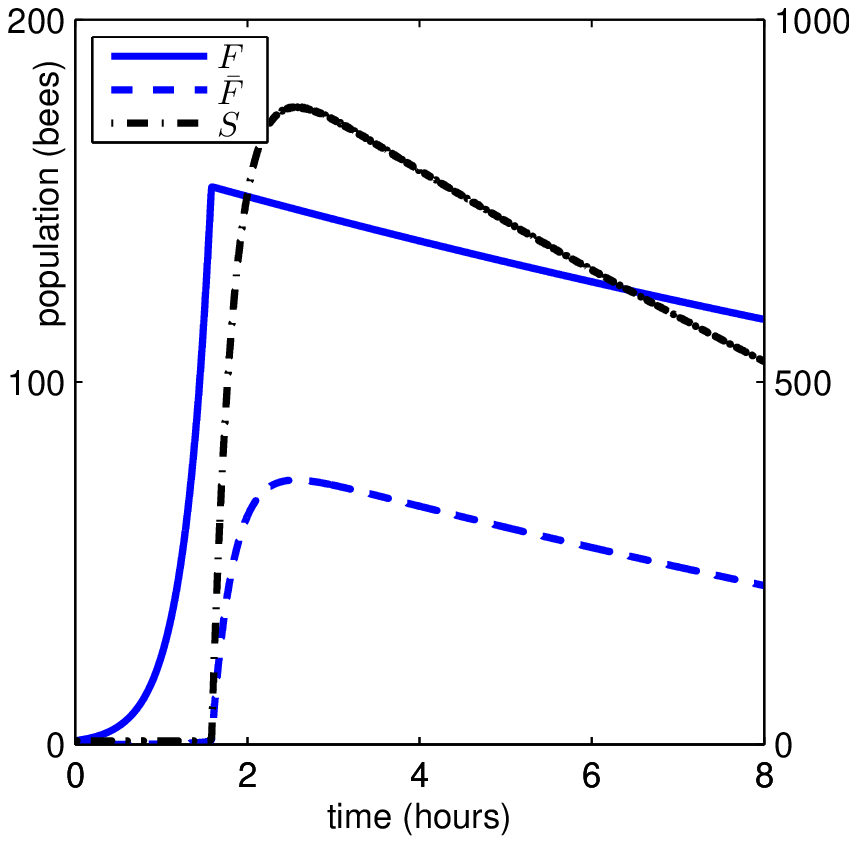}}
  \subfloat[\label{fig:timecourseHQLR}]{\includegraphics{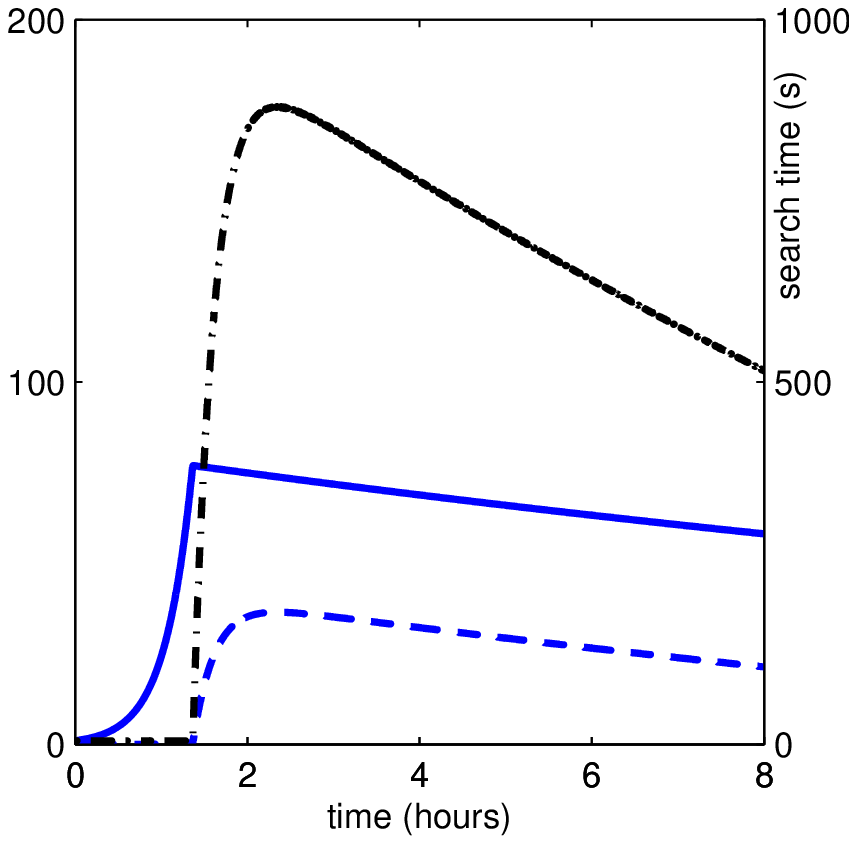}}\\
  \subfloat[\label{fig:timecourseLQHR}]{\includegraphics{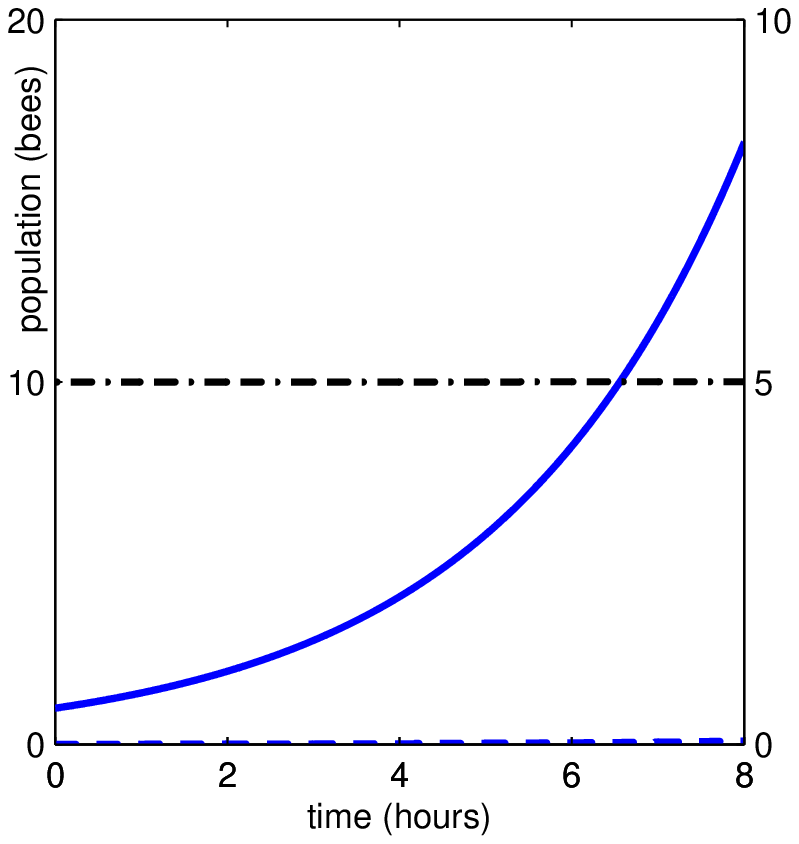}}
  \subfloat[\label{fig:timecourseLQLR}]{\includegraphics{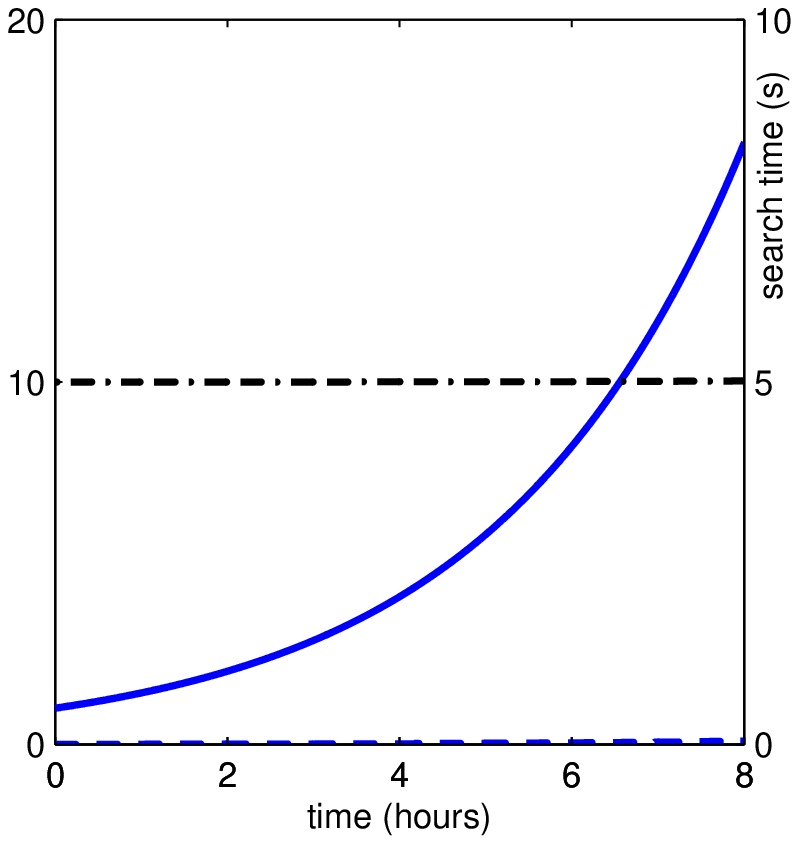}}\\
  \caption{Timecourses using the model defined by Eq. \ref{eq:system} with varying $Q$ and $R$ and other parameters as defined in Table. \ref{tab:param}. Figs. \ref{fig:timecourseHQHR} and \ref{fig:timecourseHQLR} have high quality ($Q=3.0$) environments and Figs. \ref{fig:timecourseLQHR} and \ref{fig:timecourseLQLR} have low quality ($Q=0.9$) environments. Figs. \ref{fig:timecourseHQHR} and \ref{fig:timecourseLQHR} have a large number of receivers ($R=100$) and Figs. \ref{fig:timecourseHQLR} and \ref{fig:timecourseLQLR} have a smaller number of receivers ($R=50$).  The legend is defined in Fig. \ref{fig:timecourseHQHR}. Changing $R$ dramatically affects the results of simulations using high values of $Q$, however this has no effect on simulations using low values of $Q$.}
  \label{fig:timecourse}
\end{figure}

Fig. \ref{fig:timecourse} displays the results of numerical simulations of Eq. \ref{eq:system} with the parameters of Table \ref{tab:param}. The simulated period was eight hours. This period was chosen because it conforms to the periods used in existing models \citep{camazineSneyd1,devries1998mcf,cox2003fmf}. Each of the four plots (Fig. \ref{fig:timecourseHQHR}, Fig. \ref{fig:timecourseHQLR}, Fig. \ref{fig:timecourseLQHR} and Fig. \ref{fig:timecourseLQLR}) show the population timecourses for simulations using a different combination of quality and receiver numbers. Fig. \ref{fig:timecourseHQHR} and Fig. \ref{fig:timecourseHQLR} are generated from simulations that use the same high quality environment ($Q=3$) and vary only in receiver population. This gives the simulation results in Fig. \ref{fig:timecourseHQLR}, that has half the receiver population of the simulations displayed in Fig. \ref{fig:timecourseHQHR}, producing half the forager population. The shape of the population timecourse is preserved and the search time is identical. The simulations used to produce Fig. \ref{fig:timecourseLQHR} and Fig. \ref{fig:timecourseLQLR} also differ only in receiver populations, however they both have a low quality environment ($Q=0.9$). Very different results are produced to those in Fig. \ref{fig:timecourseHQHR} and Fig. \ref{fig:timecourseHQLR}; the populations are never near equilibrium and, most significantly,  there are no differences between their population timecourses. In other words, the size of the receiver population does not affect the forager population in a low quality environment but strongly affects it in a high quality environment.

These results suggest that, in a high quality foraging environment, the population of foragers rapidly increases, as discussed in Sec. \ref{sec:steadyStates}, until the number of foragers seeking to unload becomes large relative to the number of receivers available for unloading. Once this occurs the search time increases and recruitment is inhibited, leading to much slower changes in population in accordance with Eq. \ref{eq:sysF}. The transition between these `rapid growth' and `stabilising' periods can be quite sudden. In Fig. \ref{fig:timecourseHQHR} it can be observed to occur at a cusp in the forager population timecourse just before $t=2$. This cusp typically occurs near or above the equilibrium population of $F$. In a low quality environment the rate of increase in the forager population is far slower and so the the number of foragers remains small and search time never becomes large.

This suggests that in a low quality environment the population of foragers is highly sensitive to $Q$, which controls its rate of growth. In a high quality environment the rate of growth is so fast that the onset of high search times is most important in determining the forager population. We find then that the quality, perceived by foragers, determines the population in low quality environments. When the quality is high however, it is receiver numbers that determine the forager population, since by Eq. \ref{eq:searchTime} the search time is dependent on the receiver population. That is, we can consider the quality of the foraging environment as providing positive feedback to foragers and the receiver population as providing negative feedback to foragers. The positive feedback dominates when the forager population is small and the negative feedback dominates when the forager population is large.

We offer a biological interpretation for this. A colony with large nectar reserves may need to risk only a small number of foragers in source exploitation. An efficient and simple way to do this is to limit the population of receivers, which will increase search time once this small number of foragers is active, preventing further recruitment. When the colony has depleted its reserves, for example after an extended period of rain, it can make many receivers available and foraging numbers will become constrained only by forager population size. Foragers, however, need not respond to high receiver numbers by recruiting to the maximum level if the nectar source quality is low. Thus, the two behavioural classes are interpreting different sets of data, i.e. foraging environment quality or colony nectar need, and using these different sets of information to collaboratively reach a decision about the appropriate population of foragers. The collaboration is via the search time, a value constructed in a self-organised manner through the interactions of foragers and receivers. This interpretation is biologically sensible; receivers, who store nectar, presumably understand the need of the hive for nectar and hence, by their accessibility for unloading, communicate to foragers that the hive does or does not require a greater intake. Likewise, foragers have knowledge of the foraging environment and are best suited to determining that the quality is too poor to be worth investing in more expeditions.

\begin{figure}[htbp]
  \centering
  \includegraphics{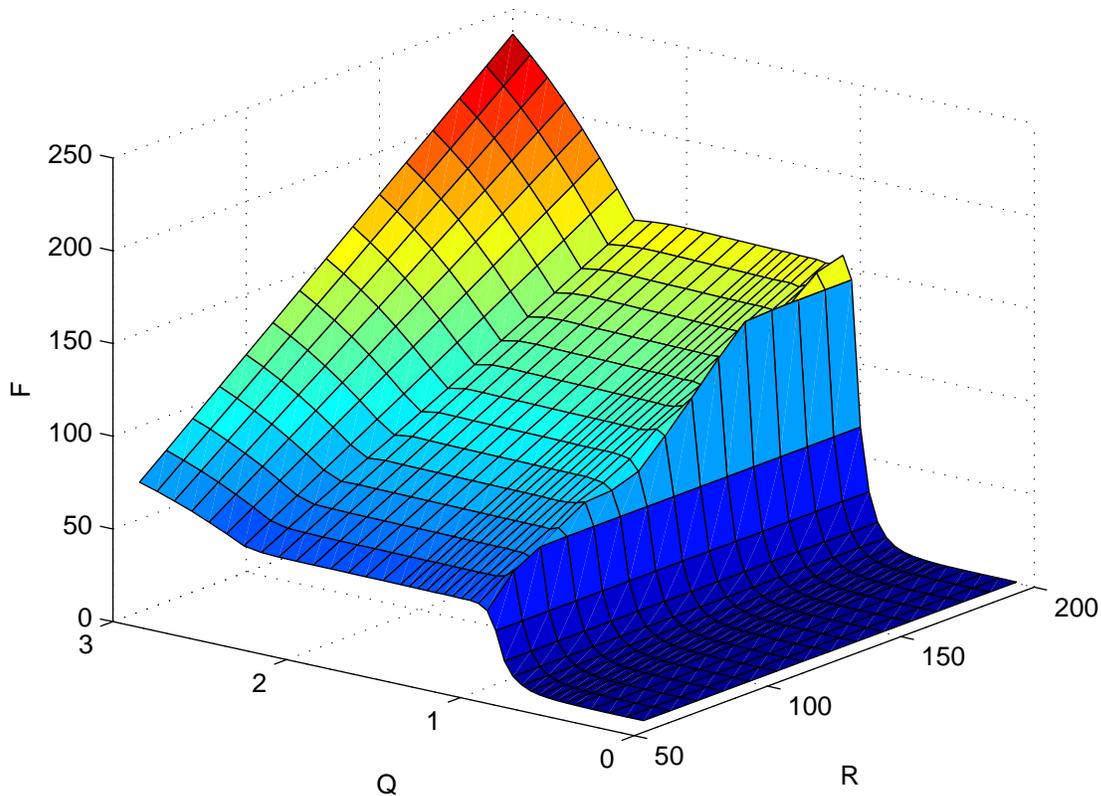}
  \caption{Forager populations after eight hours against a range of likely Q and R values. At a threshold of $Q \approx 1$ the population switches from being sensitive to $Q$ to being sensitive to $R$. The populations may exceed the stable equilibrium point temporarily. This is apparent around $Q=1$ where the graph displays a dip as $Q$ increases since simulations with quality $Q \gtrsim 1$ have decreased further from their peak after eight hours than those with a lower quality. This corresponds to the cusp that occurs as the forager numbers change in the transition from the `rapid growth' to the `stabilising' phase. A similar effect is visible for $Q>2$, where there is considerable over-shoot beyond the equilibrium population and the populations are yet to decrease to equilibrium.}
  \label{fig:FvsQvsR}
\end{figure}

A graphical representation of this is given in Fig. \ref{fig:FvsQvsR}. This figure shows the population of foragers eight hours after the commencement of foraging against a range of values for $R$ and $Q$. It is clear that, for most values, the population is sensitive to either R or Q but not to both. That is, for low qualities it is the quality of the forage source that determines the forager population and for high qualities it is the receiver population. For $Q<1$, the forager population is typically still growing after eight hours. For some values of $R$ there is a ridge at $Q \approx 1$ because the population in the simulation with a slightly higher quality has already commenced its stabilising phase and its forager population is already decreasing (c.f. Fig. \ref{fig:timecourseHQHR}). For $Q \in [1,2]$ the plot is flat along the $Q$ axis since the simulations have reached equilibrium within eight hours. For higher qualities the plot's steepness increases, reflecting those simulations whose forager populations have peaked well above equilibrium and not reached equilibrium after eight hours (c.f. Fig. \ref{fig:bar}).

\section{Extending the model with receiver recruitment via the tremble dance\label{sec:trembling}}
\begin{figure}[htbp]
 \centering
  \includegraphics{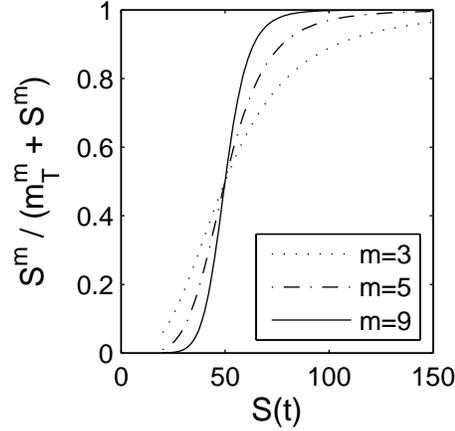}
 \caption{The effect of different values of $m$ on the trembling response. The best fitting median and coefficient values were chosen by comparing with results in \cite{seeley1992Trem}.}
\label{fig:trembleResponse}
\end{figure}

The tremble dance \citep{seeley1996hbs, dyer2002bdl} is performed by foragers experiencing high search times, and is associated with increases in the receiver population. We seek to understand how receiver recruitment affects the collective decision about forager numbers, and so we extend the simple model (Eq. \ref{eq:system}) to include tremble dancing. Let $U$ be the unit step function and $\omega$, $m$ and $m_T$ be, respectively, the maximal recruitment rate due to tremble dancing, the `steepness' of the response to tremble dancing and the minimum search time for which tremble dancing occurs. The equation that we use to model receiver population change is
\begin{equation}
 \frac{d R(t)}{d t} = U (S(t)-m_T)  \frac{\omega S(t)^m}{S(t)^m+m_T^m}.  \label{eq:receiverTrembling}
\end{equation}
Here there is no change in receiver population unless the search time is greater than $m_T$. When $S(t) > m_T$ the receiver population increases at an increasing rate as search time increases. For $\omega=0$ the system reduces to the simple model of Eq. \ref{eq:system}. The values of $m_T$ and $m$ were determined in a similar manner to $m_S, m_Q, j$ and $k$, by plotting for a range of values (Fig. \ref{fig:trembleResponse}) and comparing to \cite[Fig. 7]{seeley1992Trem}; we find that $m=5$ and $m_T=30$ are suitable values.
We begin our analysis of the new system by calculating the effect that the introduction of Eq. \ref{eq:receiverTrembling} has on the  steady state. $F(t)$ is unchanging whenever $F(t) = 0$ (trivially, though this may be unstable when the population is perturbed) or whenever 
\begin{equation}
S(t)	= S^*	 = \sqrt[k]{Q^j \frac{f_r m_S^k}{f_s m_Q^j}}. \label{eq:trembleSearch}
\end{equation}
\hide{
\begin{align*}
		& \frac{\Gamma_2 (S(t),Q_i)}{\Gamma_1 (S(t),Q_i)}	&  = \frac{f_r}{f_s}	\notag \\
\therefore	& \frac{S(t)^k m_Q^j}{m_S^k Q^j}			&  = \frac{f_r}{f_s}	\notag \\
\therefore	& S(t)						&  = \sqrt[k]{Q^j \frac{f_r m_S^k}{f_s m_Q^j}}. 
\end{align*}
}
When $S^* < m_T$, $R(t)$ eventually stabilises and then the rest of the system also stabilises. If, however, $S^* \ge m_T$ then the search time steady state $S^*$ (Eq. \ref{eq:trembleSearch}) continually induces receiver population increases and the entire system loses stability. By rearranging Eq. \ref{eq:trembleSearch} we can see that the system will be stable whenever $Q < m_Q \left((m_T^k / m_S^k) (f_s/f_r) \right)^{1/j}$.
\hide{
When $F(t)$ is stable, $\bar{F}(t)$ will become stable \note{appendix / to be proved} and so $S(t)$ is stable whenever both $F(t)$ and $\bar{R}(t)$ are. For $S(t) < m_T$, $R(t)$ is stable and hence $\bar{R}(t)$ \note{prove} is, however for $S(t) > m_T$ and $\omega \ne 0$ the population of $R(t)$ increases and the system loses stability. That is, by Eq, \ref{eq:trembleSearch}, we will have a stable forager population whenever
\begin{align*}
		& m_T^k							&  > Q^j  \frac{f_r m_S^k}{f_s m_Q^j} \\
i.e. \quad	& Q	 					&  < \sqrt[j]{m_T^k \frac{f_s m_Q^j}{f_r m_S^k}}
\end{align*}
and an unstable one otherwise. }For the standard parameters used in the simulations and defined in Table \ref{tab:param} this implies that for $Q\lesssim 3$ the forager population should reach a stable value, for higher values of $Q$ it will continually increase.

This stability criterion is independent of $R(t)$, so although high search times will increase the receiver population, this increase does not resolve the instability. Hence for high quality sources, the increases in both receivers and forager populations will continue until the colony has exhausted its populations of potential foragers or potential receivers.

This is observed in the numerical simulations presented in Fig. \ref{fig:timecourseTrembling}. Fig. \ref{fig:timecourseTrembling3} is generated using an identical configuration to Fig. \ref{fig:timecourseHQHR}, with the extra inclusion of Eq. \ref{eq:receiverTrembling} with $\omega=5$. The forager and receiver populations become unstable, continually increasing. In reality the colony would soon exhaust its population of potential foragers. A series of jagged oscillations occur for the search time when it exceeds the threshold $m_T$ defined in Eq. \ref{eq:trembleSearch}. When the search time exceeds this threshold there is a sudden influx of receivers which then causes the forager population to increase until the search time threshold is reached again. In Fig. \ref{fig:timecourseTrembling2} we have a similar set of parameters, however we have reduced the foraging quality so that $Q=2.0$, which is below the stability threshold. In this simulation, as predicted, the forager population trends towards a single value. During the rapid growth phase, receiver recruitment occurs due to the high amplitude of the search time oscillations, but this ceases as the system stabilises.\hide{ Fig. \ref{fig:timecourseTrembling1} has again been generated with the same parameters except that $Q=1.0$, which makes it comparable to the simulation used to generate Fig. \ref{fig:timecourseLQHR}. We find here that the inclusion of Eq. \ref{eq:receiverTrembling} has no effect in this low quality environment, since $S(t)$ is always less than $m_T$. This is consistent with the previous conclusions concerning low quality environments in that receiver numbers have no influence over the forager population.} A simulation (not shown) generated over the same parameter set except with $Q=0.9$ produces results identical to that in Fig. \ref{fig:timecourseLQHR}. We find here that the inclusion of Eq. \ref{eq:receiverTrembling} has no effect in this low quality environment, since $S(t)$ is always less than $m_T$. This is consistent with the previous conclusions concerning low quality environments in that receiver numbers have no influence over the forager population.
\hide{The parameter $\omega$ does not not affect the steady state but only the dynamics. Not sure this is worth mentioning.}
\begin{figure}[htbp]
 \centering
  \subfloat[\label{fig:timecourseTrembling3}]{\includegraphics{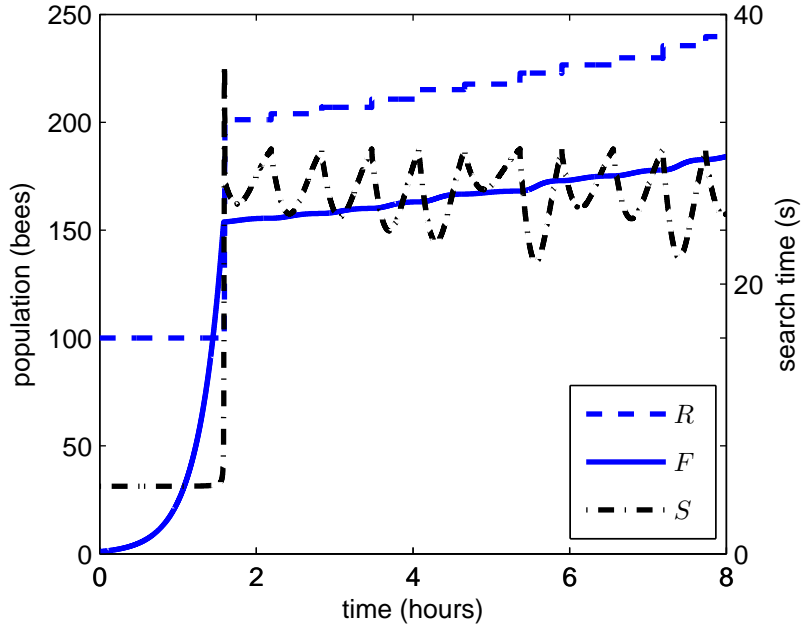}} \\
  \subfloat[\label{fig:timecourseTrembling2}]{\includegraphics{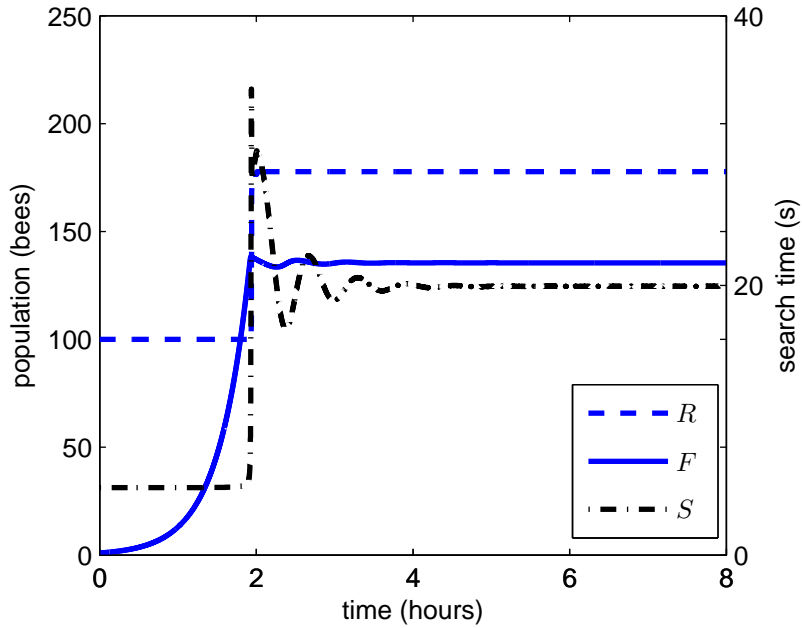}}
  \caption{Population timecourses for a model with parameters identical to Fig. \ref{fig:timecourseHQHR} except that Eq. \ref{eq:receiverTrembling} is included in the model with $\omega=5$. The dynamics are identical up to $t \approx 2$, at which point tremble dancing commences and receiver numbers begin to increase. This point coincides with the end of the rapid growth phase and commencement of the stabilising phase, discussed in Sec. \ref{sec:numerics}. In Fig. \ref{fig:timecourseTrembling3} $Q = 3$ and $S^* > m_T$, and so the receiver population increases indefinitely. The search time undergoes oscillations due to repeated cycles of receiver increase followed by forager increase causing the populations to vary widely. In Fig. \ref{fig:timecourseTrembling2} $Q = 2$ and $S^* < m_T$, and so the receiver and forager populations eventually stabilise.}
  \label{fig:timecourseTrembling}
\end{figure}
\hide{
\begin{figure}[htbp]
 \ContinuedFloat
 \centering
  \subfloat[Q = 1.\label{fig:timecourseTrembling1}]{\includegraphics{timecourseTrembling1.eps}}
\end{figure}}

These results describe a mechanism by which the equilibrium population of foragers may increase. Due to, amongst other factors, the high degree of genetic diversity in \apis\ colonies, the many stimuli thresholds at which individuals initiate action vary \citep{myerscough2004}; this is no doubt true for the tremble dancing response amongst the pool of potential receiver bees. The effect of a mechanism similar to those described by \cite{fuchsMoritz199} and \cite{Jones04} would be beneficial as the stable state populations would increase smoothly as the source quality increases. When there is a range of response thresholds, receivers play a similar role to that described in this simple model. Fundamentally, however, this model states that receivers can adapt to exceptionally high quality environments in such a way that forager numbers can increase further and thereby take full advantage of foraging conditions.

\section{Extending the model with multiple forage sources\label{sec:multipleSources}}

In the field an \apis\ colony will simultaneously exploit multiple sources, each with different quality. We therefore extend the model defined by Eq. \ref{eq:system} to include more than one nectar source. We denote the quality of $n$ sources as $Q_1, Q_2, \hdots, Q_i, \hdots, Q_n$, and add new forager equations for each source so that Eq. \ref{eq:system} is replaced by
\begin{align}
\frac{d F_i(t)}{dt} = & f_r F_i(t) \ \Gamma_1 (S(t),Q_i)  - f_s F_i(t) \  \Gamma_2 (S(t),Q_i), \notag \\
\frac{d \bar{F_i}(t)} {d t} = &  \frac{F_i(t)-\bar{F}_i(t)}{f_a} -\frac{\bar{F}_i(t)} {S(t)}, \\
\frac{d \bar{R}(t)} {d t} = & \frac{R(t) - \bar{R}(t)} {r_s} - \frac{ \sum_{i=1}^{n} \bar{F}(t)} {S(t)} \notag 
\end{align}
and search time (Eq. \ref{eq:searchTime}) is given by $S(t) = s_s \left(\sum_{i=1}^{n} \left(\bar{F}_i (t) \right) + \bar{R}(t)\right)  / \bar{R} (t) $.

We use this extended model to investigate how the inhibitory effect of high search times affects the population and the allocation of foragers among different sources, and to explore whether the results of the single source model (Eq. \ref{eq:system}) also apply to multiple source models.

\begin{figure}[htbp]
 \centering
  \subfloat[\label{fig:timecourseMultipleHQ}]{\includegraphics{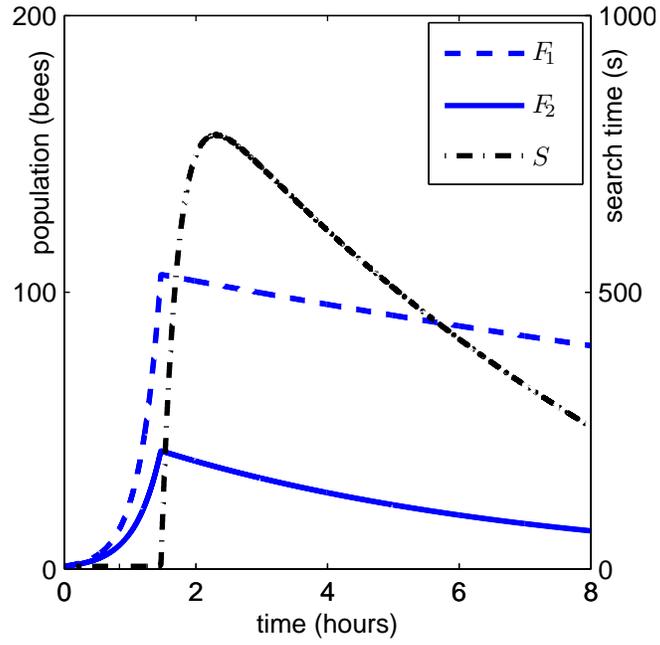}}\\
  \subfloat[\label{fig:timecourseMultipleLQ}]{\includegraphics{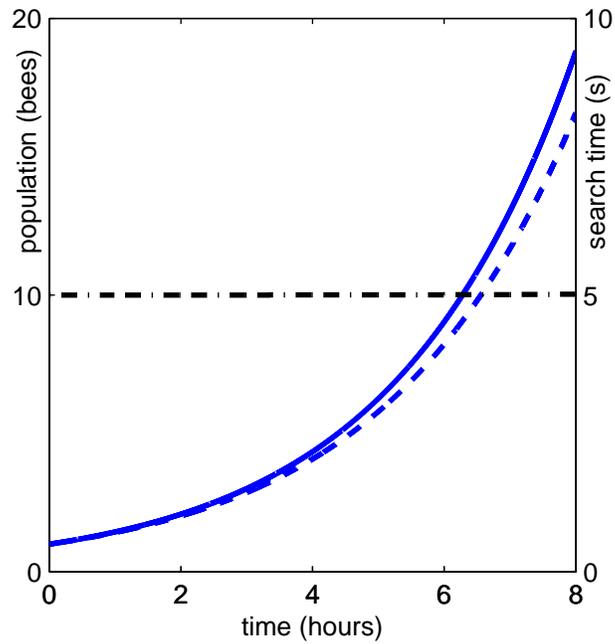}}
  \caption{Population timecourses for a model with parameters identical to those in Fig. \ref{fig:timecourse}, except that there are two forage sources. In both plots there are one hundred receivers, so they correspond to Fig. \ref{fig:timecourseHQHR} and Fig. \ref{fig:timecourseLQHR}. Fig. \ref{fig:timecourseMultipleHQ} has two high quality nectar source ($Q_1 = 3, Q_2=2$) and Fig. \ref{fig:timecourseMultipleLQ} has two low quality ones ($Q_1=0.9, Q_2=0.91$).}
  \label{fig:timecourseMultiple}
\end{figure}

The results of simulations with two nectar sources, with qualities denoted $Q_1$ and $Q_2$, are presented in Fig. \ref{fig:timecourseMultiple}. Fig. \ref{fig:timecourseMultipleHQ} displays the population timecourses from a simulation of a high quality environment, in which $Q_1=3$ and $Q_2=2$. The total population of foragers in this plot is approximately equal to the population of foragers in Fig. \ref{fig:timecourseHQHR}, which also presents results from a simulation of a high quality, though single source, environment with $Q=3$. The equilibrium populations of the two simulations are identical. The similarity between the total populations is consistent with the idea, explored previously, that the receiver population imposes an upper limit on the forager population. The simulation suggests that receivers achieve this by limiting the \textit{total population} of foragers rather than by limiting the population at each nectar source. Further, the time when the population dynamics move from a `rapid growth' phase to a `stabilising' phase is the same. Other characteristics of the figures are very similar, such as search times, available receiver population and the shape of the timecourses. For a variety of measures then, the single source quality model of Eq. \ref{eq:system} is equivalent to aggregating the population of foragers at multiple sources in high quality environments. This is expected because receivers don't know which nectar source foragers arrive from.

The distribution of foragers between the two sources changes during the rapid growth phase, with the higher quality source acquiring a larger percentage of the foragers \hide{(as expected during a phase dominated by exponential growth)}until the stabilising phase is entered, when the distribution becomes nearly static. The model assumes that there is no mechanism by which receivers direct the foragers to one particular source - they do not `manage' foraging. Instead, they specify the colony-wide nectar goal via the receiver population size and foragers determine their allocation to each source based on each source's quality. Te model predicts that, as usual, bees will display a decentralised and self-organised approach to colony management.

Fig. \ref{fig:timecourseMultipleLQ} presents the population timecourses from a simulation of a low quality environment, in which $Q_1=0.9$ and $Q_2=0.91$. The forager population at each source is equal to that of the population of foragers in Fig. \ref{fig:timecourseLQHR}, which presents results from the corresponding simulation of a single source environment with $Q=0.9$, and hence the total forager population in this plot is extremely close to twice that of the single source simulation. This is consistent with the results of the model with a single low quality source, in that foragers determine their own population on the basis of source quality without being influenced by receivers. The distribution of foragers is determined by the quality, and since $Q_1$ and $Q_2$ are nearly identical so are the populations foraging at these sources.

The numerical results of the multiple source simulations are consistent with those of the single source simulations. This suggests that the analytic results for the simple model (Sec. \ref{sec:simpleModel}) are also relevant to this multiple source model. As such, we can consider the single source model as aggregating multiple sources into one general term by thinking of $Q$ as representative of the quality of the entire environment rather than just a single nectar source.

\section{Extending the model with dynamic forage sources\hide{ and stochastic search times}\label{sec:dynamicSources}}
\begin{figure}[htbp]
 \centering
  \subfloat[\label{fig:switchEarly}]{\includegraphics{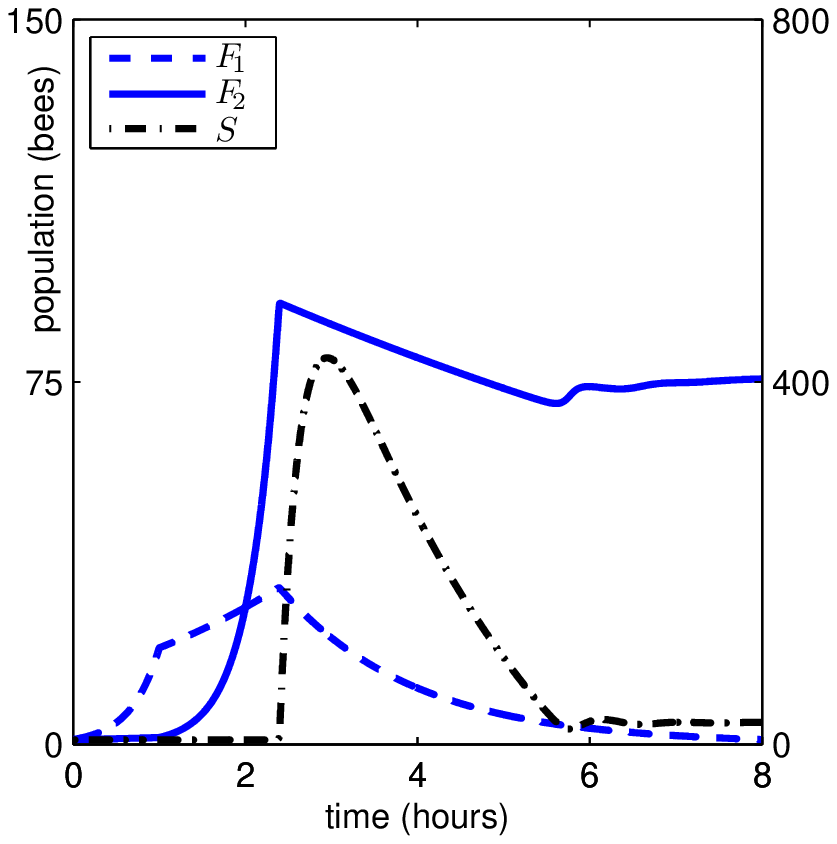}}
  \subfloat[\label{fig:switchLate}]{\includegraphics{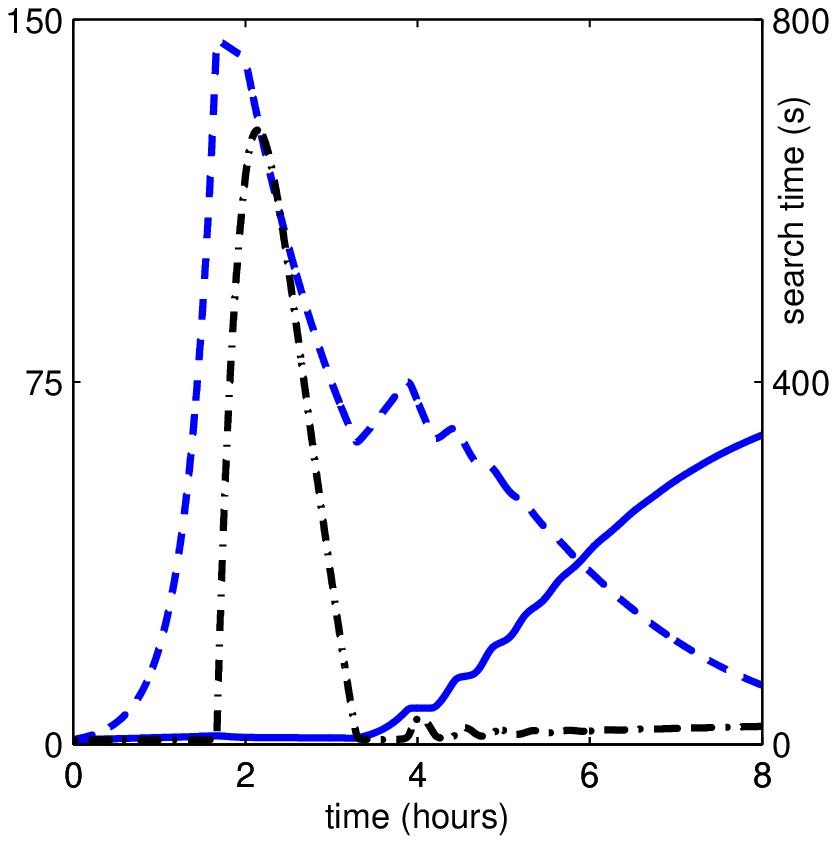}}\\
  \subfloat[\label{fig:switchEarlyTremble}]{\includegraphics{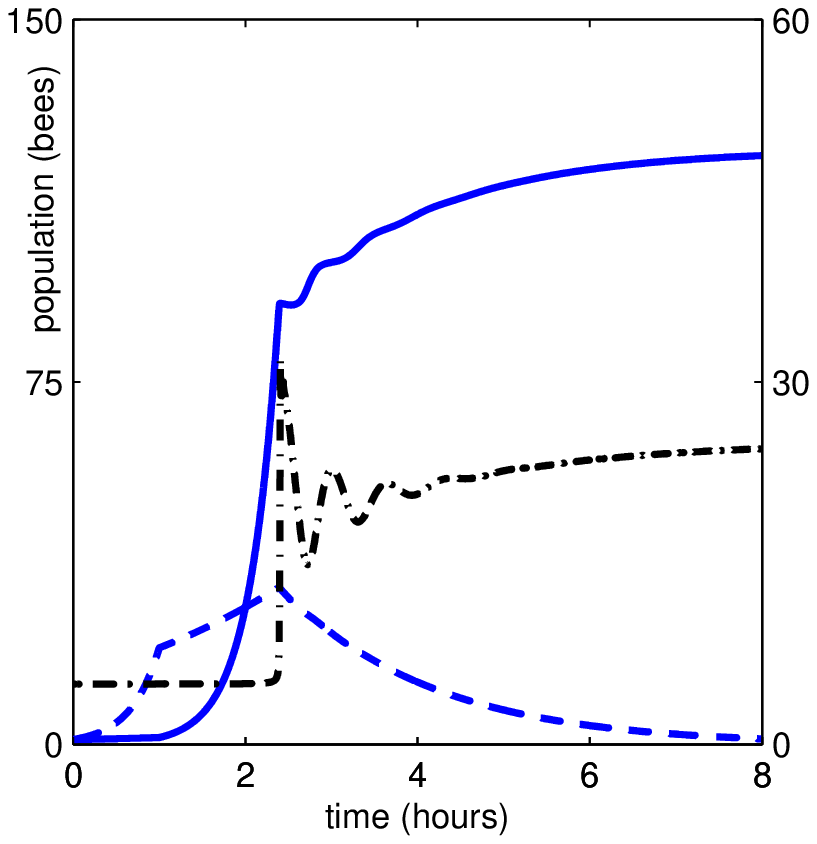}}
  \subfloat[\label{fig:switchLateTremble}]{\includegraphics{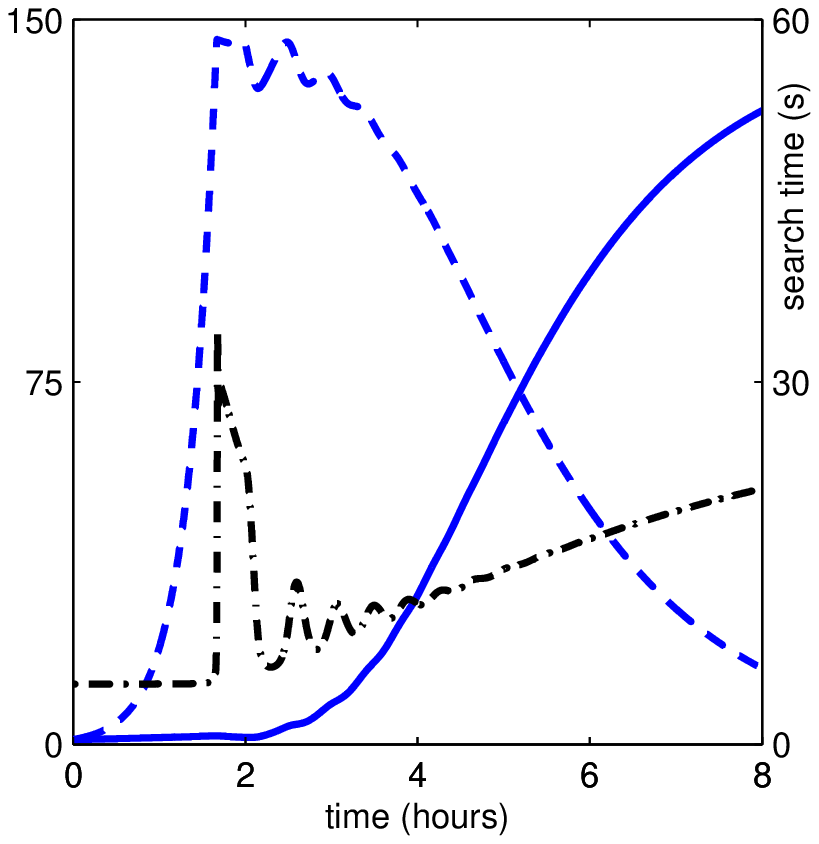}}\\
  \caption{Population timecourses for simulations of switching sources. Sources switch quality between $2.5$ and $0.9$. Parameter values are from Table \ref{tab:param} and $R=100$. In Fig. \ref{fig:switchEarly} nectar sources switch quality at $t=1$ and there is no tremble dancing. The colony adapts to the changed environment. In Fig. \ref{fig:switchLate} nectar sources switch quality at $t=2$ and there is no tremble dancing. The colony takes a long time to adapt to the changed environment. In Fig. \ref{fig:switchEarlyTremble} nectar sources switch quality at $t=1$ and there is moderate tremble dancing. The colony rapidly adapts to the changed environment. In Fig. \ref{fig:switchLateTremble} nectar sources switch quality at $t=2$ and there is moderate tremble dancing. The rate and extent of adaption is improved as compared with Fig. \ref{fig:switchLate}.}
  \label{fig:timecourseSwitch}
\end{figure}

The quality of a nectar source may change over the course of a day \citep{ginsburg1983Di}, due to changes in weather, inter-species competition levels or plant characteristics. How a colony adapts to this is crucial to its success, and so we seek to understand how search time affects forager populations when source quality changes. We follow the scenario of \cite{camazineSneydS} in which there are two nectar sources, one initially of low quality and one initially of high quality, that switch qualities after a few hours (denoted $T_s$) so that,
\begin{displaymath}
Q_1 = \left\{ \begin{array}{ll}
2.5  & \textrm{if $t<T_s$}\\
0.9 & \textrm{otherwise}
\end{array} \right.
\textrm{and} \quad
Q_2 = \left\{ \begin{array}{ll}
0.9 & \textrm{if $t<T_s$}\\
2.5  & \textrm{otherwise}
\end{array} \right.
.
\end{displaymath}
Using the simple model of Eq. \ref{eq:system}, we generate simulations of two different scenarios with this type of nectar quality switching. Each simulation initially has one hundred receivers with other parameters as defined in Table \ref{tab:param}. Population timecourses of the forager populations and search times are presented in Fig. \ref{fig:timecourseSwitch}. In Fig. \ref{fig:switchEarly} $T_s = 1$ and the foraging population rapidly changes its efforts to exploiting the better source, as with \cite{camazineSneydS}. In Fig. \ref{fig:switchLate} the parameters are identical but $T_s = 2$. The better source is not exploited in this scenario, suggesting a failure in the decision making process.

The colony's inability to adapt to the quality change in Fig. \ref{fig:switchLate} is surprising given its successful adaption to the apparently similar scenario recorded in Fig. \ref{fig:switchEarly}. The reason for the failure is that during the first simulation the switch occurred during the `rapid growth' phase, during which search times are small, and the second simulation's switch occurred during the `stabilising' phase, during which search times are high. High search times inhibit recruitment, regardless of source quality, and so during the second simulation there simply wasn't enough recruitment occurring to increase the forager population at source two. By contrast, in the first simulation there were low search times for more than two hours after the source change. Previous results consistently show that the rapid growth phase is associated with fast changes in population whereas the stabilising phase is associated with slow population changes. We note, further, that by Eq. \ref{eq:sysF}, if
$$
\frac{F_1(t)} {F_2(t)} > \frac{Q_2^j}{Q_2^j + m_Q^j} \frac{Q_1^j + m_Q^j}{Q_1^j}
$$
and $Q_1 < Q_2$, then $\frac{d F_1}{d t} > \frac{d F_2}{d t}$ and hence there is still more recruitment to the poor quality site than to the high quality one. For the parameter values used in the simulation, this means that for $\frac{F_1(t)} {F_2(t)} > 15$ (i.e. when there are $15$ times more foragers to source one than to source two) the first source will continue recruiting more foragers even after it becomes the poorer quality source. This can be observed in the timecourses of the populations in Fig. \ref{fig:switchLate} around $t = 4$. Nevertheless, the better quality site will increase its forager population at a greater proportional rate and so eventually, over a period of time much longer than a day, have the greater population as well.

With such a poor result observed in Fig. \ref{fig:switchLate}, it is natural to ask if there is something missing from the model that could improve the outcome and make it more realistic. \hide{Two phenomena seen in reality, tremble dancing and stochastic variation in search times, are likely candidates.} We will apply the same values as used to generate Fig. \ref{fig:switchEarly} and Fig. \ref{fig:switchLate} into a model extended to include tremble dancing (Eq. \ref{eq:receiverTrembling}) with $\omega = 0.1$. The resulting population timecourses are displayed in Fig. \ref{fig:switchEarlyTremble} and Fig. \ref{fig:switchLateTremble}. The results of Fig. \ref{fig:switchEarlyTremble} are similar to those in Fig. \ref{fig:switchEarly}, though with the increased populations expected due to the inclusion of tremble dancing. The results of Fig. \ref{fig:switchLateTremble}, however, improve considerably upon those in Fig. \ref{fig:switchLate}.
The improvement occurs because tremble dancing relieves some inhibitory pressure on forager recruitment, as the increased receiver numbers temporarily decrease the search time. Increasing numbers of these new recruited foragers exploit the higher quality nectar source.\hide{ The inclusion of variable search times should also mean that, even when the average search time is high, occasionally a forager will experience a short search time and recruit. In conjunction, these effects greatly improve the outcome.}

Even with the modifications, the change in forager allocation when the quality switch occurs at $T_s=2$ still appears to be slow. There are two possible reasons for this. Firstly, the model could be showing a result that is qualitatively accurate, yet its outcome is much better in practice due to behaviours that are not modelled. Secondly, the model could be quantitatively correct and this slow response is advantageous in someway to the colony, representing perhaps a collective memory effect that highly values nectar sources that have previously proven to be of good quality. It could also be seen as a part of the colony's risk-reduction strategy, as the colony allocates its foragers sub-optimally (in terms of maximising nectar influx) so that it can robustly protect against another sudden loss of the better source.

\section{Discussion}
We have presented a class-oriented model of \apis\ nectar foraging whose principal novelty is that it incorporates the receiver population. Numerical and analytic results derived from the model all suggest that the size of the receiver population determines the equilibrium population of active foragers. In general, foragers exploiting a high quality nectar source rapidly recruit more new foragers, but only while there are receivers available to quickly unload them. Once the demand for receivers significantly exceeds supply, forager recruitment is inhibited and their population stabilises.

Under the simple model defined and analysed in Sec. \ref{sec:simpleModel} and Sec. \ref{sec:numerics}, receiver numbers are fixed. Once this fixed population is reached no more bees will become receivers (though until it is reached some of the receivers may be engaged in other worker tasks). The model is modified in Sec. \ref{sec:trembling} so that new receivers are recruited by tremble dancing. The results of the simulations produced by a model that includes tremble dancing suggest that when an extremely high quality nectar source is found, the colony may indefinitely recruit new receivers, at least until the worker population is exhausted. Normally however, tremble dancing allows the colony to increase its population of receivers to a greater, but still limited, size if this is justified by the discovery of a high quality nectar source.

Another system is also plausible; some worker bees are dedicated receivers, others are available to receive under certain criteria and others are never available. Such a system allows a staggered response to tremble dancing, so that individual bees may have different thresholds \citep{Jones04} at which they will switch to receiving; some will switch as soon as they perceive a tremble dance, others only when the colony need is great. The actual scheme is of great importance in determining the role of receivers. If workers always switch roles to become receivers upon noticing a tremble dance, then they are only passive participants in the decision making process. If, however, they are influenced by their environment as well as the tremble dance then they have an important role to play in deciding how many foragers should be active (as we have modelled here). The effect on the decision making process is immense, and warrants further study. Under such a scheme, the tremble dance recruits non-receivers to act as receivers but this will only occur amongst those bees who perceive a high colony need. Once again, this makes biological sense; bees that are available to receive should be engaged in other housekeeping roles until such time as there is demand for their labour.

The results reported in Sec \ref{sec:multipleSources} suggest that receivers are not directly involved in allocating foragers to specific forage sources. Instead, their population implicitly specifies a target forager population and the foragers determine the distribution amongst different sources. We note, however, that there is experimental evidence that receivers can compare the quality of nectar coming in from different sources and preferentially respond to foragers carrying the highest quality nectar \citep{scheiner2004Su,martinez2008Su}. These foragers will have a shorter search time (and a shorter unloading time). Receivers may influence the allocation with this mechanism, since recruitment to high quality sources will be increased and recruitment to low quality sources decreased as foragers returning from high quality sources have a longer search time.

Sec. \ref{sec:dynamicSources} models a colony adapting to a dramatic change in nectar source quality. Interpreted qualitatively, the colony response depends on whether the change in quality occurs before or after the start of the `stabilising' phase - i.e. when the demand for receivers outstrips the supply. The colony smoothly adapts to a change before this phase, but takes much longer to respond correctly when the change occurs afterwards, when recruitment is inhibited by high search times. The inclusion of tremble dancing improves the response significantly. This is interesting because it highlights how flexibility within the colony, in this case in the receiver population, leads to improved outcomes.
\hide{The model results pertain to cross inhibition \ci\ between forager populations at different sources. This model hence provides an explanation of cross inhibition \note{explain} \citep{seeley:wh} %page ,142%
[also Boch 1956 in german and probably another seeley paper see seeley 86 and free 63] that does not rely upon limits to the unemployed forager population.}
\subsection{Experimental implications}

Nectar receivers have received less attention than foragers. However, our results indicate that the nectar receiver population has a significant influence on the colony's exploitation of forage sources. For instance, the model suggests that the daily foraging cycle may consist of periods in which the nectar receiver population inhibits forager population change - even as conditions themselves change. It would be interesting to see whether this inhibition does in fact occur in real hives.

In addition, the population of active nectar receivers may fluctuate due to colony need and the level of tremble dancing. The model results suggest that receivers should respond to tremble dancing only when the foraging environment is excellent and/or the colony has low nectar stores. Observations are needed to demonstrate that the activity of the nectar receivers indicates the internal state of the colony and that this, in turn, affects the forager population.
\hide{Receivers have not attracted the same level of attention as foragers. Our results suggest that they deserve more. It would be especially interesting to learn the criteria worker bees have for becoming receivers; are some worker bees ready to become receivers upon perceiving a tremble dance or are they all potential receivers just waiting for some threshold involving current nectar stores, consumption rates and tremble dancing to be met? More data concerning the responsiveness of receivers to different types of nectar would be also useful in determining how much of a role receivers play in forager distribution.

More information about the recruitment of receivers following a tremble dance would also be valuable. The equation that models it here, Eq. \ref{eq:receiverTrembling}, depends especially on the the $\omega$ parameter which is an entirely unknown value.}
\subsection{Model limitations and future extensions}
A number of extensions are suggested by the biology and by the model results. Over long periods a hive may fill up with nectar. Until new honeycomb is constructed this has the effect of increasing storage time; storage time is represented in this model by the constant parameter $r_s$. We see from Eq. \ref{eq:stableApprox} that as $r_s$ increases, the steady state populations decrease. The model could be extended to make $r_s$ vary with the level of nectar stored within the hive. This may result in greater nectar stores leading to smaller foraging levels, and exploring this in the model may elucidate the system further. In addition, including the tendency of workers to construct new honeycomb as the hive fills and being able to visualise its effect on the forager and receiver populations would be useful and interesting. 

Another extension is to allow receivers to respond more favourably to nectar gathered from particular locations. We have discussed previously how this could lead to increased foraging populations at the favoured nectar sources, but modelling it would quantify this and allow us to better understand its potential effect under different circumstances.

One further extension would be to include inspector bees (foragers who repeatedly check a previously good quality nectar source to ascertain whether it has returned to being good quality) and different subclasses of foragers that respond differently to different values of $Q$ and $S(t)$. This should be particularly relevant when studying task allocation, models of which often rely on different individual response thresholds \citep{beshers2001Thr,myerscough2004}.

\subsection{Foraging as a collective decision}

Our model treats foraging as a problem solved by a collective group intelligence; the colony analyses information and makes a decision about forager allocation. In the model two classes of bee, each with access to different, yet critical, pieces of data interact while unloading nectar in such a way that a self-organised value emerges in the form of the search time. This value provides both positive and negative feedback, depending on the current situation, and so the colony is able to make a wise choice when it decides on its forager allocation. The inclusion of receivers in this model provides an inhibition mechanism that is common in decision making models \citep{sumpter2006rpc} and which counters the activation provided by forager recruitment.
\bibliography{refs}
\end{document}